# Towards Public Administration Research Based on Interpretable Machine Learning

Zhanyu Liu [a], Yang Yu [b, *]

[a] *School of International Relations and Public Affairs, Fudan University, Shanghai, 200433, China*

[b] *School of Public Administration, Hunan University, Changsha, 410012, China*

**Abstract:** Causal relationships play a pivotal role in research within the field of public administration. Ensuring reliable causal inference requires validating the predictability of these relationships, which is a crucial precondition. However, prediction has not garnered adequate attention within the realm of quantitative research in public administration and the broader social sciences. The advent of interpretable machine learning presents a significant opportunity to integrate prediction into quantitative research conducted in public administration. This article delves into the fundamental principles of interpretable machine learning while also examining its current applications in social science research. Building upon this foundation, the article further expounds upon the implementation process of interpretable machine learning, encompassing key aspects such as dataset construction, model training, model evaluation, and model interpretation. Lastly, the article explores the disciplinary value of interpretable machine learning within the field of public administration, highlighting its potential to enhance the generalization of inference, facilitate the selection of optimal explanations for phenomena, stimulate the construction of theoretical hypotheses, and provide a platform for the translation of knowledge. As a complement to traditional causal inference methods, interpretable machine learning ushers in a new era of credibility in quantitative research within the realm of public administration.

## 1. Introduction

Causal relationships constitute a pivotal subject within the realm of public administration research. It refers to the situation where a change or presence of one factor leads to corresponding changes in another factor (Imbens and Rubin, 2015). They are regarded as a central tenet of social science research. The examination of causal relationships not only fulfills the human quest for understanding social phenomena but also furnishes a more scientific, precise, and dependable framework for addressing societal issues. As a critical branch of the social sciences, public administration places a premium on causal reasoning (Frank et al., 2023). Numerous discussions within this field fundamentally revolve around causal inference, including assessments of policy impacts, rationales for selecting pilot projects, and the dynamics of governmental responsiveness. Delving into causal relationships in public administration research aids in uncovering the consequences and mechanisms of managerial decisions, thereby supplying empirical evidence to enhance governmental efficacy and the delivery of more impactful public services (James et al., 2017). Consequently, comprehending the derivation of credible causal inference from the intricate fabric of reality is a paramount concern for researchers in public administration.

A necessary step to achieve credible causal inference is incorporating "prediction" into public administration research. To derive credible causal inference, numerous causal inference methods from economics have been introduced into public administration studies (Dague and Lahey, 2019). These methods fall into two main categories: first, randomized controlled trial, which involves artificially assigning interventions randomly to create experimental data for testing causal relationships (George et al., 2020). Second, quasi-experimental design, which uses statistical methods such as matching, difference-in-differences, and instrumental variables to identify causal relationships from observational data. These methods effectively address issues like reverse causality and omitted variables in causal inference (Dague and Lahey, 2019). However, they do not completely guarantee the authenticity of derived causal relationships. Another necessary condition for causal relationships to hold is predictability (Watts, 2014). Simply put, if a causal relationship exists, it should be able to predict the outcomes of unknown phenomena. This aspect has not received sufficient attention in empirical research in public administration and the broader social sciences (Hofman et al., 2017). As we know, empirical research in public administration largely relies on parameter estimation, such as the commonly used ordinary least squares regression. These methods evaluate causal relationships primarily based on the significance of regression coefficients, rather than the out-of-sample predictive performance of the fitted models (Molina and Garip, 2019; Mullainathan and Spiess, 2017). Given this situation, integrating "prediction" into the knowledge framework of quantitative public administration research is crucial (Hofman et al., 2021).

The development of machine learning offers an opportunity for social science research to achieve prediction (Lundberg et al., 2022a). Specifically, machine learning-based modeling has two main advantages. On the one hand, algorithms used in machine learning modeling, such as support vector machines, neural networks, and random forests, can more accurately uncover patterns in data, including nonlinear relationships and high-dimensional interactions. This reduces the fitting bias of machine learning algorithms. On the other hand, machine learning employs robust model validation techniques, such as hold-out validation and k-fold cross-validation, to monitor and prevent overfitting to data noise (Athey and Imbens, 2019). These advantages collectively improve the predictive performance of machine learning models (Geman et al., 1992).

It is worth noting that for a long time, machine learning predictions were not used for causal identification. This is because the internal mechanisms by which machine learning models predict outcome variables are often highly complex (Hofman et al., 2021). In machine learning models, the influence of independent variables or predictive features on outcome variables cannot be easily expressed through regression coefficients, as in linear models commonly used in social sciences. In other words, traditional machine learning models can generally answer what the prediction result is but not provide insights into causal relationships. This feature of traditional machine learning models has raised concerns because it could lead to ethical issues such as freedom, bias, and so on (Rudin et al., 2022). Against this backdrop, interpretable machine learning has been proposed as a complementary approach. Interpretable machine learning refers to methods that make the decision-making processes of predictive models easily understandable by humans (Molnar, 2020). Its core idea is to use post hoc interpretive techniques to present the roles of different features in complex models based on the "input-output" relationships (Molnar, 2020). Recent studies have suggested that interpretable machine learning can be used for causal inference (Basu et al., 2018). For example, interpretable machine learning can first ensure the predictive performance of relationships between variables, and then use traditional causal inference methods to further validate these relationships (W James Murdoch et al., 2019). Thus, interpretable machine learning could potentially be applied to public administration research to complement existing causal inference methods and address the "predictability" issue in causal inference.

Based on the above discussion, this paper will explore the application prospects of interpretable machine learning in public administration. It should be noted that this study is not a systematic methodological review, as interpretable machine learning has not yet been widely adopted in public administration research. Instead, this paper provides a preliminary exploration of the concept of interpretable machine learning, its applications in social science research, implementation steps, and disciplinary value. Specifically, the remainder of the paper is organized as follows: Section 2 introduces the concept of interpretable machine learning and its applications in social sciences; Section 3 elaborates on the implementation steps of interpretable machine learning; Section 4 discusses the disciplinary value of interpretable machine learning for public administration; and Section 5 examines the limitations of interpretable machine learning and concludes the study.

## 2. Understanding interpretable machine learning

### 2.1 Machine learning and interpretable machine learning

Machine learning is a branch of artificial intelligence that aims to enable computers to automatically discover patterns from data, thereby accomplishing tasks such as classification, regression, dimensionality reduction, and information extraction. These tasks can be categorized into two main branches: supervised learning and unsupervised learning. Supervised learning typically seeks to fit a structure to a dataset to predict outcome variables, whereas unsupervised learning usually attempts to extract features from a dataset (Athey, 2019). This study focuses on prediction problems related to supervised learning. The typical process of supervised learning can generally be divided into three steps. First, dataset construction, which includes defining the outcome variable and selecting and extracting predictive features. Second, model training, which refers to the process of using preprocessed data to train predictive models. This involves selecting appropriate machine learning algorithms, tuning algorithm parameters, and fitting the model. Third, model evaluation, which involves using an independent test dataset to assess the predictive performance of the trained model. This step may also include comparing the performance of models trained with different algorithms to select the optimal model (Valizade et al., 2022). Once model evaluation is completed, researchers can select a relatively optimal predictive model. However, in most studies, the best predictive models often rely on complex algorithms, such as ensemble methods and neural networks, where the parameters do not clearly show the relationships learned by the model (Rudin, 2019). In this case, the predictive model becomes a "black-box model" (Suvorova, 2021), making the relationships between predictive features and outcome variables difficult to understand.

The concept of interpretable machine learning was introduced to address the "black-box" issue of traditional machine learning models (Molnar, 2020). It should be noted that interpretable machine learning is not a substitute for traditional machine learning but rather a supplement. The primary goal of traditional machine learning is to achieve accurate predictions by building high-performance models, while the goal of interpretable machine learning is to provide explanations of the model's prediction process and decision-making while maintaining a certain level of predictive performance (Suvorova, 2021). Compared to traditional machine learning, interpretable machine learning introduces an additional step of model interpretation. The purpose of this step is to use post hoc interpretation techniques to visualize the relationships between predictive features and outcome variables. This primarily includes two aspects. First, feature importance evaluation, which uses importance evaluation techniques such as feature importance based on tree models (Lundberg et al., 2018), permutation importance (Altmann et al., 2010), and SHAP values (Lundberg and Lee, 2017) to evaluate the contribution or importance of different features to the model's predictive outcomes. Second, feature effect visualization, which employs visualization techniques such as partial dependence (PD) plots, accumulated local effects (ALE) plots, and LIME (Apley and Zhu, 2020) to show how changes in features affect the predicted outcome and the direction of these changes.

## 2.2 The application of interpretable machine learning in social science

Currently, interpretable machine learning has been applied in a small amount of social science research. Based on the domain of the dependent variables, these studies can be categorized into four types:

The first type pertains to public safety studies. Such studies aim to predict events that threaten public safety, such as terrorism and crime, and to explain the impact of various features on the likelihood or frequency of these events. For example, Python et al.(2021) trained a prediction model for terrorist events using XGBoost and visualized the relationships between population density, economic level, and event probability using ALE plots. X. Zhang et al.(2022) trained a prediction model for regional crime levels using XGBoost and assessed feature importance with SHAP values. The results showed that the proportion of non-local residents and individuals aged 25-44 were the most significant predictors of regional crime levels. Kim and Lee (2023) trained a prediction model for regional crime density using LightGBM and evaluated feature importance with SHAP values. The results indicated that the number of retail stores, population, and points of interest density were critical predictors of regional crime density. Ahmed et al.(2023) trained the New Zealand road accident dataset using random forests and other methods, and conducted importance assessment with SHAP values. The study shows that road category and the number of vehicles involved in accidents are important predictive features.

The second type focuses on economic markets studies. These studies develop prediction models for objectives such as consumer behavior, corporate management, and market prices, and analyze key predictive features. For instance, Chen et al.(2021) trained a prediction model for consumers' purchase of travel services using a stacked model and visualized the relationship between service ratings and purchase probability using PD plots. Z. Zhang et al.(2022) trained a prediction model for corporate financial distress using LightGBM and evaluated feature importance with SHAP values. Operating profit growth rate was identified as the most critical predictor of financial distress. Jabeur et al.(2021) trained a prediction model for oil price crashes using random forests and assessed feature importance with SHAP values. The results revealed that crude oil prices were the most significant predictors of oil price crashes. Chen et al.(2021b) trained customer information, operational behavior data, and online comments using various machine learning algorithms, and then conducted feature importance assessment with SHAP values and Partial Dependence Plots. The results show that ratings are an important feature affecting online travel purchases.

The third type relates to social family studies. These studies focus on predicting family or household demands or developmental statuses and explaining key influencing factors. For example, Xenochristou et al.(2021) trained a prediction model for household water demand using random forests and evaluated importance using permutation importance. The results showed that household occupancy was found to be the most critical predictor of water demand. Ledesma et al.(2020) trained a prediction model for household wealth indices using random forests and evaluated feature importance with SHAP values. The results showed that regional night-time light intensity, the proportion of 4G users, and the proportion of public schools with potable water were identified as the most important predictors of household wealth indices. Paes et al.(2023) trained a prediction model for residents' income levels using neural networks and assessed feature importance with SHAP values. The results showed that educational level and aging were identified as significant predictors of income levels. Liu and Hu (2024) trained a prediction model for crowdfunding performance in environmental projects using GBDT and evaluated feature importance with SHAP values. The results showed that the fundraising target was identified as the most crucial predictor of performance. Yang et al.(2024) trained random forests on crowdsourced data collected from Strava and conducted importance assessment with SHAP. The results indicated that green space area and the green view Index are significant features affecting running. Zou et al.(2025) trained a prediction model on Guangzhou travel survey data using a random forest and conducted importance assessment with SHAP values. The results showed that the start time of activities and activity sequence dependency factors are important characteristics of home-based activities on weekdays.

The final type pertains to intelligence information studies. These studies apply interpretable machine learning to classic topics in bibliometrics and informatics, such as paper citations and social media text classification. For example, Beranová et al.(2022) trained a prediction model for paper citation levels using neural networks and presented the effects of different predictive features on a representative sample using LIME plots. Ha (2022) trained a prediction model for paper citation levels using CatBoost and assessed feature importance with SHAP values. The results indicated that publication source was the most critical predictor of paper citation levels. Adarsh et al.(2023) trained a prediction model for depressive tweets in media using SVM+KNN and visualized the effects of specific feature words on a representative tweet using LIME plots.

An overview of these studies is provided in Table 1. The above analysis indicates that interpretable machine learning has begun to gain traction in social science research, indirectly proving its applicability. However, it is notable that this approach has yet to be adopted in public administration research. This highlights the necessity of promoting its application in public administration studies.

**Table 1**

The Application of Interpretable Machine Learning in Social Science Research

| Affiliated Fields | Outcome Variables | Predictors | Optimal Algorithm | Interpretation Technique | References |
|---|---|---|---|---|---|
| Public Safety | Terrorism Events | Population; Transportation; Remoteness; Economy; Welfare; Minerals; Political System; Terrorist Organizations, etc. | XGBoost | ALE Plot | (Apley and Zhu, 2020) |
| | Regional Crime Density | Population; Buildings; Roads; Points of Interest; Civil Disputes; Urban Landscape | LightGBM | SHAP Values | (Kim and Lee, 2023) |
| | Regional Crime | Population; Wealth; Roads; Points of Interest | XGBoost | SHAP Values | (X. Zhang et |

| | | Level | | | al., 2022) |
|---|---|---|---|---|---|
| Economic Markets | Online Travel | Consumer Demographics; Historical Order Features; Behavioral Operations | Stacking | SHAP Values; PD Plot | (Chen et al., 2021a) |
| | Corporate Financial Distress | Solvency; Structural Ratios; Operational Capability; Profitability; Growth Capability; Per-Share Ratios; Cash Flow | LightGBM | SHAP Values; PD Plot | (Z. Zhang et al., 2022) |
| | Oil Price Collapse Events | Green Energy; Global Environmental Index; Stock Market | XGBoost | SHAP Values | (Jabeur et al., 2021) |
| Social Family | Household Poverty Index | Social Media Ads; Remote Sensing Data; Points of Interest | Random Forest | SHAP Values | (Ledesma et al., 2020) |
| | Environmental Crowdfunding Performance | Fundraising Goals; Target Domain; Fundraising Duration; Project Description; Characteristics of Fundraising Organizations; Characteristics of Project Leaders | GBDT | SHAP Values; ALE Plot | (Liu and Hu, 2024) |
| | Household Water Demand | Historical Consumption; Time; Weather; Household Characteristics | Random Forest | ALE Plot; Permutation Importance | (Xenochristou et al., 2021) |
| | Resident Income Level | Demographics; Internet Infrastructure; Network Type; Digital Exposure | Neural Network | SHAP Values | (Paes et al., 2023) |
| Intelligence Information | Paper Citation Level | Language; Type; Open Source; Source; Discipline; Number of References; Number of Authors; Number of Countries; Number of Institutions, etc. | CatBoost | SHAP Values | (Ha, 2022) |
| | Paper Citation Level | Number of Authors; Impact Factor; Journal; JCR Category; Abstract Word Vectors | Neural Network | SHAP Values; LIME | (Beranová et al., 2022) |
| | Media Depression Tweets | Frequency of Depression-Related Words | SVM+KNN | LIME | (Adarsh et al., 2023) |

*Source of information: Author's own creation.*

## 3. Implementation of interpretable machine learning

To provide researchers with a deeper understanding of the practical application of interpretable machine learning, this section elaborates on its implementation steps in detail. Specifically, the discussion will cover four aspects: dataset construction, model training, model evaluation, and model interpretation. Fig. 1 illustrates the implementation process of interpretable machine learning.

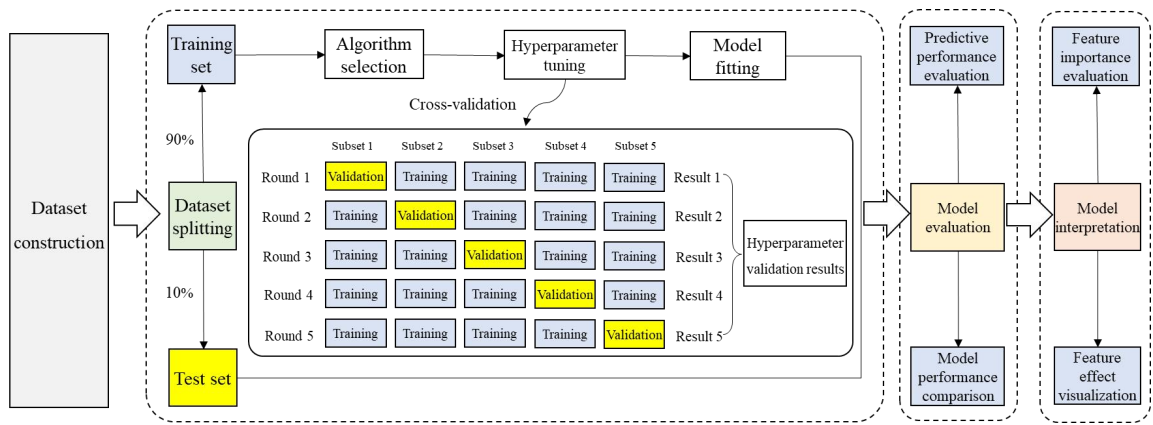

**Fig. 1.** The Implementation Process of Interpretable Machine Learning

*Source of information: Author's own creation.*

## 3.1 Dataset construction

### 3.1.1. Outcome variable

The first step in constructing a dataset is defining the outcome variable for the predictive model (also known as the target variable, response variable, or label). This process is akin to the measurement of dependent variables in traditional causal inference. The definition of the outcome variable typically involves two substeps: first, measuring or operationalizing the concept. The outcome variable generally pertains to a specific concept. To make an abstract objective suitable for training a predictive model, it must be converted into concrete data indicators. For instance, in public administration research, predicting "efforts in building government reputation" requires a representation in data form, such as using the number of related articles on government portal websites as a proxy.

Second, defining the type of predictive problem. Predictive problems can be divided into two types: regression tasks and classification tasks. The former is used for predicting continuous variables, while the latter is suited for categorical variables. However, researchers sometimes convert continuous variables into categorical variables to leverage powerful classification algorithms (Bibi et al., 2008; Halawani et al., 2011). Therefore, unlike traditional causal inference, constructing a dataset in machine learning involves an additional step of defining the type of predictive problem. For instance, in public administration research, when predicting "public support for policies", measured using a 7-point scale, researchers can choose either to treat the variable as a continuous one for regression or convert it into a binary variable for classification. In fact, such transformations are occasionally employed in traditional causal inference studies, but they are usually conducted as a preliminary step for robustness checks or supplementary analysis (Hu et al., 2022).

*3.1.2. Predictive feature*

The second step in constructing a dataset is extracting and processing predictive features. This step is similar to the measurement of independent variables in traditional causal inference. It also consists of two parts: first, selecting and extracting predictive features. Feature selection in interpretable machine learning differs from traditional machine learning. Its feature selection needs to be based on causal relationships or theoretical drivers, rather than data availability (Hofman et al., 2021). Otherwise, even if important variable relationships are obtained through model interpretation techniques, researchers will not be able to understand the causal mechanisms and theoretical implications behind those relationships. However, interpretable machine learning can refer to the feature extraction approach in traditional machine learning, using advanced techniques to extract theoretically meaningful features from unstructured data such as text, images, and so on (Athey and Imbens, 2019; Cranmer and Desmarais, 2017; Grimmer et al., 2021).

Second, preprocessing feature data. To achieve better model training results, the predictive feature data needs to be preprocessed before being applied to model training. This typically includes four categories: (1) One-hot encoding. Categorical variables generally have no natural order or ranking relationship. Using integer encoding (e.g., 0, 1, 2) may mislead the model and cause it to incorrectly learn the order relationship between categorical variables. One-hot encoding avoids this preference or ordering issue by expanding each value into a binary feature. (2) Handling outliers. The presence of outliers can cause the model to overfit or overgeneralize to these anomalies. This may lead to reduced model performance on normal data. Outliers are typically defined as data points greater than "mean + 3×standard deviation" or less than "mean - 3×standard deviation[1]." (3) Logarithmic transformation. For skewed data distributions, applying a natural logarithmic transformation can reduce the scale differences in the data, making it smoother and more symmetrical (Curran-Everett, 2018). (4) Normalization. Different features usually have different scales and ranges, which may cause some features to have a greater impact on the model. By normalizing and scaling the features to the same range, the impact weight of different feature on the model can be more balanced (Singh and Singh, 2020).

---

[1] Assuming that the data follows a normal distribution, the probability of a value 3 standard deviations away from the mean is 0.003, which is a minimal probability event.

## *3.2 Model training*

### *3.2.1. Dataset splitting*

The first step in model training is to split the dataset into a training set and a test set. The purpose of this is to reserve part of the data to evaluate the generalization ability of the machine learning model on unseen data. The typical ratio for the training set and test set is 9:1, meaning 90% of the samples from the dataset are randomly selected as the training set for model fitting, and the remaining 10% are reserved as the test set for model prediction.

It is important to note that when there is an imbalance between positive and negative samples in a classification task dataset, it is necessary to first use resampling techniques to balance the positive and negative samples in the training set before model training (Neunhoeffer and Sternberg, 2019). Directly using an imbalanced dataset for model training may lead to overfitting of the majority class samples and underfitting of the minority class samples (Mohammed et al., 2020). For example, suppose 90% of the samples in the dataset are assigned a value of 1 and 10% are assigned a value of 0. A model that simply predicts all samples as 1 might achieve a high accuracy. However, such a model would likely perform poorly in predicting negative class samples. The issue of dataset imbalance can be addressed using resampling techniques, which include oversampling and undersampling. Generally, oversampling techniques tend to have better optimization effects on the model than undersampling techniques (Mohammed et al., 2020).

### *3.2.2. Algorithms and parameters*

The second step in model training is selecting an algorithm and tuning its hyperparameters. First, supervised machine learning can be implemented using various algorithms, including linear regression, decision trees, support vector machines, random forests, extreme gradient boosting, neural networks, deep learning, and so on (Athey and Imbens, 2019). The choice of algorithm to fit the model depends on several factors. For causal inference purposes, algorithm selection can be considered from two aspects: (1) Dataset size. For large datasets, complex models like neural networks and deep learning may help uncover intricate patterns. For small datasets, simpler models such as linear models or traditional machine learning algorithms like decision trees might suffice. If the dataset size is mismatched with the algorithm complexity, model prediction performance may be limited due to overfitting or underfitting. (2) Explainability requirements. If the researcher only wants to explain the linear effect of predictive features on the outcome variable, traditional linear models are more suitable. If the researcher aims to further explain the nonlinear effects of predictive features on the outcome variable and the interactions between predictive features, ensemble algorithms based on decision trees are a better choice.

Second, in machine learning, hyperparameter tuning aims to find the optimal hyperparameter configuration to improve the model's performance and generalization ability. Key points for achieving algorithm tuning include: first, understanding what the hyperparameters of the used algorithm are and their functions. Hyperparameters are parameters that need to be manually set before model training, such as learning rate, regularization coefficient, depth of decision trees, etc. By understanding the meaning and impact of hyperparameters, they can be better adjusted to improve model performance. Second, determining the value range or possible values for each hyperparameter. This can be guided by experience, domain knowledge, literature research, or experiments. Make sure the search space is wide enough to cover the potential optimal hyperparameter configuration. Third, choosing a hyperparameter tuning method. Common tuning methods include grid search and random search. Grid search explores specified hyperparameter combinations, while random search randomly selects combinations from the given hyperparameter space for evaluation (Bergstra and Bengio, 2012). Generally, random search is more efficient and requires fewer computational resources than grid search (Yu and Zhu, 2020). Finally, evaluating and selecting the optimal hyperparameter configuration. The performance of different hyperparameter configurations is typically evaluated using cross-validation (Lones, 2021). Cross-validation divides the dataset into multiple subsets and repeatedly trains and evaluates the model to reliably estimate its performance. Cross-validation helps determine the generalization ability of the model under different hyperparameter configurations. After selecting the best hyperparameters, the algorithm can be used to train the predictive model on the training set.

## 3.3 Model evaluation

### 3.3.1. Predictive performance evaluation

The first step in model evaluation is to assess the model's predictive performance on the test set. For regression tasks, commonly used performance evaluation metrics include Mean Absolute Error (MAE), Mean Squared Error (MSE), Root Mean Squared Error (RMSE), and R-squared ($R^2$). Among these, MAE, MSE, and RMSE measure the model's prediction errors, with smaller values being better. R-squared represents the proportion of variance in the target variable explained by the model, ranging from 0 to 1. The closer the value is to 1, the better the model's predictive power.

For classification tasks, common performance evaluation metrics include Accuracy, Precision, Recall, F1 Score, and Area Under the ROC Curve (AUC). Accuracy is one of the most commonly used metrics in classification models, indicating the proportion of samples correctly predicted by the model. The higher the accuracy, the more consistent the model's predictions are with the true labels. Precision measures the accuracy of the model in predicting positive class samples. It represents the proportion of correctly predicted positive class samples among all samples predicted as positive. Recall measures the proportion of actual positive class samples that are correctly predicted as positive by the model. It represents the proportion of correctly predicted positive class samples among all actual positive class samples. The F1 score is the harmonic mean of Precision and Recall, taking both accuracy and recall into account. It is a comprehensive evaluation metric suitable for imbalanced data situations. The AUC value is calculated by ranking the samples according to their predicted probability, calculating the true positive rate and false positive rate (1-true negative rate) at different thresholds, plotting the ROC curve, and computing the area under the curve.

In real classification tasks, AUC and F1 score are the most commonly used evaluation metrics because these metrics reflect the overall performance of the model better (Chicco and Jurman, 2020; Halimu et al., 2019). Both metrics have a range of 0 to 1. Typically, values of 0.5 indicate random predictions, values greater than 0.7 indicate acceptable model performance, and values greater than 0.8 indicate good predictive performance (Hosmer et al., 2013).

### 3.3.2. Model performance comparison

The second step in model evaluation is to compare the performance of different models. During model training, researchers often train multiple predictive models based on different algorithms or hyperparameters. By evaluating multiple models, we can select the one with the highest accuracy, lowest error, or other relevant metrics as the best model, and proceed with further tasks such as model interpretation (Henninger et al., 2023). A common approach for comparing model performance is to directly compare the evaluation scores of multiple models on the reserved test set. However, some scholars suggest incorporating statistical tests to provide more reliable and statistically meaningful comparison results (Lones, 2021). Specifically, researchers can use resampling techniques to randomly sample with replacement from the original dataset to generate multiple datasets. By evaluating the models' prediction performance on multiple datasets, the distribution of performance metrics can be obtained. A more robust comparison of model performance can then be made by conducting t-tests or Mann-Whitney U tests on the performance metric distributions of different models[2].

## *3.4 Model interpretation*

### *3.4.1. Feature importance evaluation*

The first step in model interpretation is to assess the importance of predictive features. Evaluating feature importance helps identify which features play a decisive role in the prediction results. Feature importance evaluation refers to determining which features contribute the most or are the most influential in a machine learning model. As mentioned earlier, there are three common methods for assessing feature importance: tree-based feature importance (Lundberg et al., 2018), permutation importance (Altmann et al., 2010), and SHAP values (Lundberg and Lee, 2017). However, this study recommends using the latter two for feature importance evaluation, as tree-based feature importance tends to overestimate the importance of continuous features and underestimate the importance of root-level features (Lundberg et al., 2018; Zhou and Hooker, 2020).

SHAP values (SHapley Additive exPlanations) reflect the marginal contribution of a feature to the prediction results after being added to the model. This technique treats the model prediction process as a cooperative game, where the features are the players, and the prediction results are the game's payoff. Calculating SHAP values involves evaluating all possible feature permutations to determine the marginal contribution of features within these permutations (Lundberg and Lee, 2017). An example of SHAP values is shown in Fig. 2 (left). The wider the range of a feature's SHAP values, the higher its relative importance. Permutation importance evaluates the impact of a feature on model prediction by disrupting the relationship between the feature and the true outcome variable. Specifically, this technique first uses the original data for model prediction to establish a baseline performance metric; then, it randomly permutes the data of a specific feature and calculates the performance metric after permutation. Finally, it compares the difference between the permuted performance metric and the baseline metric—the larger the difference, the greater the feature's contribution to model performance (Altmann et al., 2010). An example of permutation importance is shown in Fig. 2 (right).

---

[2] Both the T-test and the Mann-Whitney U test can be used to compare the means, the former for normally distributed data and the latter for non-normally distributed data.

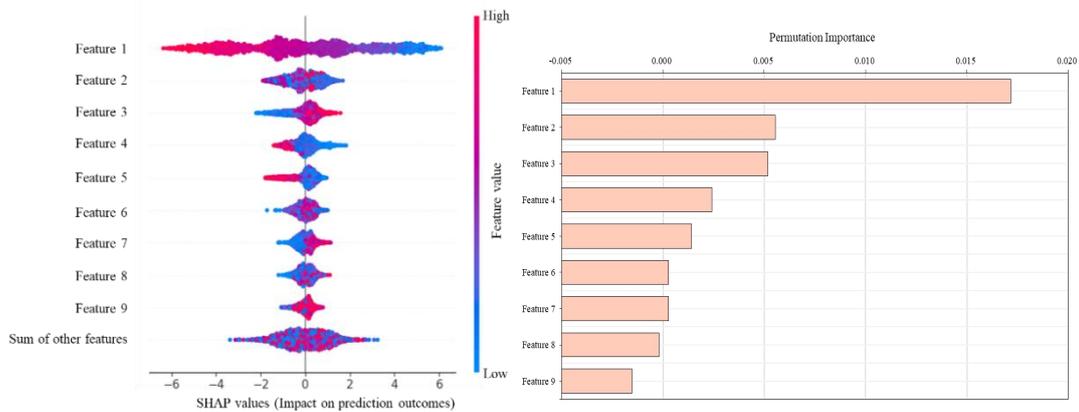

**Fig. 2.** Example of SHAP values for predictive features (left) and Permutation Importance (right)

*Source of information: Author's own creation.*

*3.4.2. Feature effect visualization*

The second step in model interpretation is to visualize the effects of predictive features on the outcome variable. As mentioned earlier, Partial Dependence (PD) plots, Accumulated Local Effects (ALE) plots, and LIME can all be used to illustrate the effects of predictive features on the outcome variable (Apley and Zhu, 2020). This study focuses on discussing ALE plots in detail, as they have clear advantages over the other two techniques. On the one hand, LIME is a local interpretation method that focuses on explaining the prediction results of individual samples, whereas ALE plots provide insights into the impact of features on prediction results at the dataset level. On the other hand, while PD plots can also offer explanations at the dataset level, they require features to be mutually independent. In contrast, ALE plots can provide accurate model interpretations even when features are interdependent (Lucas, 2020).

ALE plots come in two forms: first-order ALE plots and second-order ALE plots. The former visualizes the effect of a single feature on the prediction results using a curve, while the latter uses a heatmap to visualize the interaction effects between two features. For clarity, this paper provides examples of both first-order and second-order ALE plots, as shown in Fig. 3.

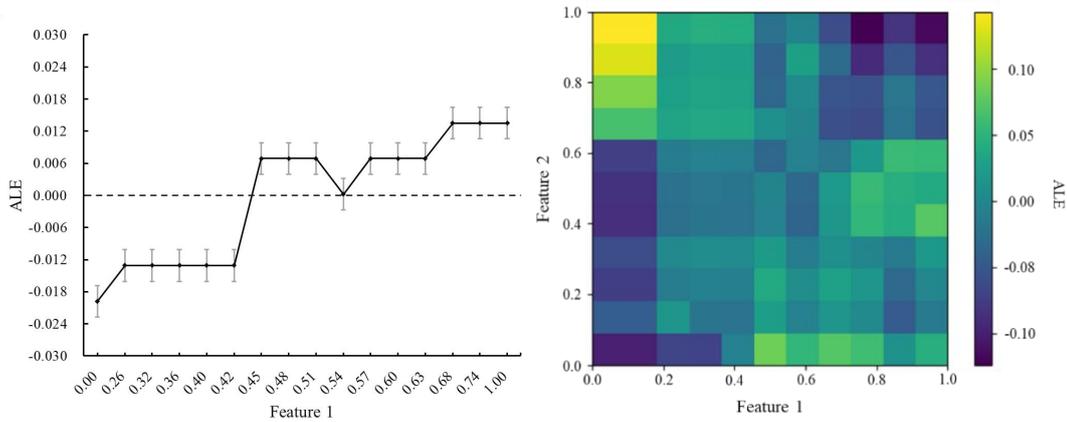

**Fig. 3.** Example of first-order ALE plot (left) and second-order ALE plot (right)
Source of information: Author's own creation.

In a first-order ALE plot, the horizontal axis represents the range of feature values, while the vertical axis indicates the local effect magnitude of the feature at corresponding values. Local effects can be understood as the difference between the prediction result at a specific feature value and the average prediction level (Molnar, 2020). For example, in the example plot (left), the second point represents that when the value of Feature 1 is 0.26, the prediction result is 0.012 lower than the average level. Additionally, first-order ALE plots provide a 95% confidence interval for the local effects. By examining the shape of the curve in the first-order ALE plot, researchers can determine the nature of the relationship between the predictive feature and the outcome variable.

In a second-order ALE plot, each cell in the heatmap represents the interaction effect of different value combinations of two features. The brighter the color of a cell, the stronger the interaction effect of that combination. It is important to note that interaction effects are not merely the simple sum of the main effects of the features; rather, they represent the additional effects beyond the main effects. This aligns with the well-known systems theory concept that "the whole is greater than the sum of its parts." For example, in the example plot (right), the top-left cell has an ALE value greater than 0.1. This indicates that when Feature 1 takes a value between 0 and 0.1 and Feature 2 takes a value between 0.9 and 1.0, their combined effect exceeds the sum of their main effects by more than 0.1. By analyzing the heat distribution in a second-order ALE plot, researchers can identify which predictive features exhibit interaction effects.

## 4. Disciplinary value of interpretable machine learning for public administration

By summarizing the concepts and implementation processes of interpretable machine learning, it becomes evident that this approach differs significantly from traditional causal inference based on statistical analysis. These distinctions span multiple stages, including data, modeling, evaluation, and interpretation (see Table 2). It is apparent that interpretable machine learning could potentially serve as a new paradigm for causal inference. This section will explore the disciplinary value of this paradigm for public administration, highlighting its prospects in enhancing inferential generalizability, selecting optimal explanations for phenomena, inspiring theoretical hypothesis construction, and providing a vehicle for knowledge translation.

**Table 2.**

Comparison of Interpretable Machine Learning and Traditional Causal Inference

| Steps | Causal Inference Paradigm | |
| --- | --- | --- |
|  | Traditional Causal Inference Based on Statistical Analysis | Interpretable Machine Learning |
| Data | <u>Variable Measurement</u><br>• Measure the dependent variable<br>• Measure the independent variables | <u>Dataset construction</u><br>• Define the outcome variable<br>• Extract and process predictive features |
| Modeling | <u>Model Specification</u><br>• Specify the model structure | <u>Model Training</u><br>• Dataset splitting<br>• Select algorithms and hyperparameters |
| Evaluation | <u>Model Evaluation</u><br>• Assess goodness of fit | <u>Model Evaluation</u><br>• Evaluate predictive performance<br>• Compare model performance |
| Interpretation | <u>Empirical Analysis</u><br>• Interpret regression coefficients | <u>Model Interpretation</u><br>• Assess feature importance<br>• Analyze feature effects |

Note: The underline represents the corresponding names of each step in the two paradigms.

*Source of information: Author's own creation.*

### *4.1 Enhancing the generalizability of public administration inference*

Firstly, interpretable machine learning helps to enhance the generalizability of public administration inference. With the widespread development of causal inference methods, causal inference in public administration have been increasing. However, these inferences are often not entirely convincing to the audience. People often worry that these inferences stem from researchers "searching for statistical significance" or "P-value manipulation" (Brodeur et al., 2020; Hofman et al., 2017). Seemingly reasonable inference are often just the result of researchers trying to justify their explanations and fit them to the data outcomes (Hofman et al., 2021). To change this status quo, the key lies in enhancing the generalizability of causal inference. A reliable causal inference should be universally applicable, maintaining a certain level of accuracy even in different populations or contexts (W. James Murdoch et al., 2019). To achieve this, public administration scholars have made some efforts. For instance, some scholars advocate for replication studies to enhance the generalizability of public administration inference. However, replication studies remain limited to experimental research and have not been fully extended to observational studies (Pedersen and Stritch, 2018; Walker et al., 2017). Moreover, the cost-related issues of replication studies have not been properly addressed. These problems can be solved through the interpretable machine learning approach proposed in this paper. This method provides an intuitive metric for the generalizability of causal inference in observational studies—out-of-sample prediction performance (Daoud and Johansson, 2024). This may further enhance the generalizability of public administration inference.

## *4.2 Exploring the optimal explanation of public administration phenomena*

Secondly, interpretable machine learning helps to select the optimal explanation for public administration phenomena. As we know, explanatory research in public administration is experiencing explosive growth. This means that there may be multiple explanations behind the same public administration phenomenon. For example, for the classic phenomenon of policy innovation diffusion, existing research has proposed dozens of potential influencing factors (Mallinson, 2021, 2020). While these numerous influencing factors help enrich the knowledge base, practitioners may be more concerned with selecting a few of the most important factors (Boon et al., 2024). In other words, the excessive growth of theoretical explanations for public administration phenomena urgently requires "downsizing" to select the optimal explanation from many alternatives (Leavitt et al., 2021). Interpretable machine learning has good prospects in achieving this goal, as predictive ability is a good method for comparing the strengths of different explanations for the same phenomenon (Cranmer and Desmarais, 2017). According to existing studies, there are two main ways to use interpretable machine learning to select the optimal explanation: first, by training different models based on different theoretical frameworks and comparing the prediction performance of the models to select the best framework. For instance, existing studies have constructed predictive models of civil war outbreaks based on different theoretical frameworks to assess the strengths and weaknesses of the theories (Blair and Sambanis, 2020). Second, by placing alternative explanatory factors in the same model, and using interpretive techniques such as feature importance to select the most important factors. For example, existing research has included numerous influencing factors on voter behavior in the same model to filter out important variables (Kim et al., 2020).

## *4.3 Stimulating the construction of public administration theory*

Thirdly, interpretable machine learning helps to inspire the construction of hypotheses in public administration theory. In the past, the innovation and development of public administration theory mostly relied on prior knowledge and scholars' academic imagination. While this model ensures the normative nature of theoretical innovation, it also brings two limitations: On one hand, theoretical hypotheses tend to oversimplify. For instance, most theoretical models proposed by scholars assume that relationships between variables are linear (Shrestha et al., 2021), because complex nonlinear relationships or interactions are often difficult to explain or memorize. But we must acknowledge that these relationships exist in reality. On the other hand, scholars tend to propose grand theories rather than local theories. For example, few scholars are willing to spend time developing a public administration theory specifically for a certain functional department, while they are more inclined to develop a theory applicable to all government departments, as developing local theories is considered a less cost-effective theoretical endeavor (Lavelle-hill et al., 2021). However, for managers in specific departments, the former is often more attractive. Interpretable machine learning can address both of these issues, inspiring scholars to construct more complex public administration theoretical hypotheses: first, interpretable machine learning usually does not rely on explicit assumptions or prior knowledge but instead extracts patterns and rules from the data in a data-driven manner. Many flexible algorithms, such as decision trees, automatically account for nonlinear relationships and interactions between variables (Lavelle-hill et al., 2021; Lundberg et al., 2022b;

Shrestha et al., 2021; Shu and Ye, 2023a; Srour and Karkoulian, 2022). For researchers, using model interpretation techniques to present these complex relationships can inspire them to build relatively complex theoretical hypotheses. Second, the rules mined from the data by interpretable machine learning will carry the contextual characteristics of the data. In other words, as long as researchers use data from a specific department to train the model, the model will learn the theoretical rules applicable to that department's context (Leavitt et al., 2021). This indirectly reduces the cost for researchers to develop local theories and, to some extent, enhances their motivation to develop such theories.

### *4.4 Providing the transformation platform of public administration achievements*

Finally, interpretable machine learning helps provide platform for translating public administration achievements. One of the missions of public administration research is to provide scientific evidence to improve public administration practices. However, as a member of the social sciences, the outcome translation rate of public administration research, especially quantitative research, still has significant room for improvement. The outcome translation rate can be reflected by how much the research results improve public administration practices. If the research results provide new insights, methods, or tools that improve the efficiency, effectiveness, or public satisfaction of public administration, the outcome translation rate is higher. In the era of big data, one tool that is continuously empowering public administration practices is predictive modeling. For instance, governments use predictive models to improve resource allocation, identify high-risk populations for poverty, improve refugee resettlement efficiency, and more (Bansak et al., 2018; Kleinberg and Ludwig, 2016; Mullainathan and Spiess, 2017). A core step in interpretable machine learning is training predictive models. Although these predictive models may not directly be used for public decision-making, their theoretically driven algorithmic frameworks have the potential to guide and optimize existing public decision-making algorithms. In other words, the predictive models trained by interpretable machine learning can serve as a platform for the translation of public administration research outcomes.

## 5. Discussion and conclusion

Obtaining credible causal inference is an important goal in quantitative public administration research. To achieve this goal, quantitative research in public administration has undergone a "credibility revolution." During this revolution, attempts to explore causality in public administration have incorporated the rich causal inference statistical methods from economics, addressing issues such as omitted variables and reverse causality that lead to endogeneity problems. However, a necessary condition for causal relationships—predictability—has not received the attention it deserves. This means that "prediction" needs to be integrated into quantitative public administration research. Interpretable machine learning, an emerging method that balances "predictability" and "interpretability," has gradually been applied to social science research. This paper argues that this method can fill the gap of "prediction" in quantitative public administration research and bring about a new "credibility revolution."

Based on the above discussion, this paper explores the application prospects of interpretable machine learning in public administration. Specifically, the paper first introduces the concept of interpretable machine learning and provides an overview of its application in social science research areas such as public safety, economic markets, social families, and information intelligence. Second, the paper presents the implementation process of interpretable machine learning, covering the steps of dataset construction, model training, model evaluation, and model interpretation. Finally, the paper discusses the differences between interpretable machine learning and traditional causal inference methods, and highlights the value of this method for the field of public administration, focusing on enhancing the generalizability of inferences, selecting the optimal explanation for phenomena, inspiring the construction of theoretical hypotheses, and providing a vehicle for the transformation of research results.

It should be emphasized that interpretable machine learning is not intended to replace traditional causal inference methods, but to complement them (Shu and Ye, 2023b). While "predictability" is a necessary condition for causal relationships, it is not a sufficient condition. Just because a predictive model can accurately predict an outcome does not mean there is a true causal relationship. The relationship learned by the predictive model may be driven by correlations in the data and does not necessarily indicate the presence of a causal relationship (Valizade et al., 2022; Varian, 2014). Causality requires more in-depth causal inference methods to prove its existence by addressing potential endogeneity (Chou et al., 2023). As mentioned above, future causal inference processes could combine interpretable machine learning with traditional causal inference methods (W James Murdoch et al., 2019).